\documentclass{article}
\usepackage{spconf, amsmath,graphicx,amssymb}
\usepackage{float}
\usepackage[normalem]{ulem}
\usepackage{footnote}
\usepackage{dsfont}
\usepackage{algpseudocode,algorithm}
\usepackage{multirow}

\newif \ifwebcolor
\webcolortrue
\def \ba {\begin{array}}
	\def \ea {\end{array}}
\def \benu {\begin{enumerate}}
	\def \eenu {\end{enumerate}}
\def \bdes {\begin{description}}
	\def \edes {\end{description}}

\def \bitem {\begin{itemize}}
	\def \eitem {\end{itemize}}

\def \bml {\begin{flushleft}}
	\def \efl {\end{flushleft}}
\def \bmr {\begin{flushright}}
	\def \efr {\end{flushright}}

\def \beq {\begin{equation}}
	\def \eeq {\end{equation}}
\def \bqa {\begin{eqnarray}}
	\def \eqa {\end{eqnarray}}
\def \bqa* {\begin{eqnarray*}}
	\def \eqa* {\end{eqnarray*}}

\def \bal {\begin{align}}
	\def \eal {\end{align}}

\title{Sound Source Separation Using Latent Variational \\ Block-Wise Disentanglement}
\name{Karim Helwani$^\ast$, Masahito Togami$^\ast$, Paris Smaragdis$^{\ast, \sharp}$, and Michael M. Goodwin$^\ast$\thanks{© 2024 IEEE. Personal use of this material is permitted. Permission from IEEE must be obtained for all other uses, in any current or future media, including reprinting/republishing this material for advertising or promotional purposes, creating new collective works, for resale or redistribution to servers or lists, or reuse of any copyrighted component of this work in other works.}}
\address{$^\ast$Amazon Web Services, $^\sharp$University of Illinois at Urbana-Champaign}

\begin{document}
\ninept
\maketitle

\begin{sloppy}
\begin{abstract}

While neural network approaches have made significant strides in resolving classical signal processing problems, it is often the case that hybrid approaches that draw insight from both signal processing and neural networks produce more complete solutions.  In this paper, we present a hybrid classical digital signal processing/deep neural network (DSP/DNN) approach to source separation (SS) highlighting the theoretical link between variational autoencoder and classical approaches to SS.  We propose a system that transforms the single channel under-determined SS task to an equivalent multichannel over-determined SS problem in a properly designed latent space. The separation task in the latent space is treated as finding a variational block-wise disentangled representation of the mixture. We show empirically, that the design choices and the variational formulation of the task at hand motivated by the classical signal processing theoretical results lead to robustness to unseen out-of-distribution data and reduction of the overfitting risk. To address the resulting permutation issue we explicitly incorporate a novel differentiable permutation loss function and augment the model with a memory mechanism to keep track of the statistics of the individual sources.
  \end{abstract}
\begin{keywords}
		Source separation, Out-of-distribution robustness.
\end{keywords}
	%


\vspace{-.5cm}
\section{Introduction}
\label{sec:intro}

Real-life audio communication signals often capture a complex acoustic scene where multiple sound sources are active simultaneously. The task of segregating the sound sources in an audio stream has interested the research community for the last few decades and the recent development of source separation systems based on deep neural network (DNN) architectures refueled the hope of the community in developing reliable real-time systems for this objective. A pioneering work using DNN architectures for this so-called cocktail party scenario is the Deep Clustering approach presented in \cite{hershey2016deep}  where the presented architecture is trained to produce spectrogram embeddings that are discriminative for partition labels given only the spectrogram of the training data. While this approach was offline, more recent approaches to near real-time source separation  operate online \cite{Luo_2019} or in a block-online (segment-wise) manner \cite{li2022skim}. The general scheme of most of the state-of-the-art models follows the encoder, masker, decoder structure e.g., \cite{Luo_2019, tzinis2020sudo, subakan2021attention, luo2020dualpath}. The input to these models is the raw time-domain signal and the task of the encoder is to find a representation where the sources are separable via a masking operation. Finally, the decoder reconstructs the individual signals in isolation. The classical line of thought in approaches to blind source separation (BSS), see e.g., \cite{trini, kim2010real, 5229304}, sets relatively strict {\it a priori} assumptions about the stochastic model of the separated signals. While classical approaches are principled and often formulated to work in an online manner, the assumptions made about the demixing system and the source statistics are often limiting factors for these approaches to work in adverse conditions. On the other hand, DNN-based approaches achieve impressive performance compared to classical approaches on several datasets, however, the models are often complex and designed as end-to-end systems, such that there is limited understanding of the latent representation derived by these models and their ability to generalize. 

In the present study, we introduce a DNN model inspired by classical signal processing approaches to perform online source separation. We establish the link between classical source separation approaches and variational auto-encoders, and show how to modify the auto-encoder architecture to systematically perform the disentanglement of the sources. 
The concept of using a variational auto-encoder as a generative spectrogram model for a single clean speech source for speech enhancement and separation has been explored in previous research \cite{Bando_2018, Leglaive_2018, pariente2019statistically, li2019fast, vaemss} and for multichannel source separation in \cite{9506855} where it has been used for inference of independent factors in an unsupervised manner assuming a Gaussian latent space. In this study, we are instead interested in formulating the separation task variationally rather than approximating individual speech spectrogram densities. More specifically, we are interested in finding a transformation that decomposes the microphone signal latent representation into mutually independent multivariate components following a predetermined stochastic mixture model. The motivation of this formulation is to make the model more robust to scenarios that were unseen during training and hence to avoid overfitting.
The network accounts for the permutation problem by introducing an explicit differential permutation loss over sources and frequencies and the incorporation of a memory mechanism.%
\vspace{-.2cm}
\section{Classical Blind Source Separation}
The task of source separation in general for any number of channels is to adaptively learn a function $\mathbf{y}(n) = \mathcal{W}(\mathbf{x}(n), \boldsymbol{\theta})$ where $\mathbf{x}(n)$ represents a frame of the observable sensor signals at time index $n$, $\boldsymbol{\theta}$ are the separation model parameters, $\mathbf{x}^{\mathrm{T}}(n)=\left[\mathbf{x}_1^{\mathrm{T}}(n), \ldots, \mathbf{x}_Q^{\mathrm{T}}(n)\right]$, and $\mathbf{y}^{\mathrm{T}}(n)=\left[\mathbf{y}_1^{\mathrm{T}}(n), \ldots, \mathbf{y}_P^{\mathrm{T}}(n)\right]$, with $\mathbf{x}_q^{\mathrm{T}}(n)=\left[x_q(n), \ldots, x_q(n- L+1)\right]$ and $\mathbf{y}_p^{\mathrm{T}}(n)=\left[y_p(n), \ldots, y_p(n-D+1)\right].$
The number of sources and sensors are $P$ and $Q$ respectively, and $L$ input frame length required to estimate one sample, and $D$ the time lags defining the dimensionality of the source model. A general optimization cost function for BSS has been introduced in \cite{trini}. Although the cost function was originally presented for the determined multichannel linear convolutive case, it can be generalized to cover the under-determined and non-linear case with a few modifications. The cost function is given by
	\begin{align}\label{eq:trini}\vspace{-.1cm}
		\mathcal{J}(m, \boldsymbol{\theta})=&-\sum_{i=0}^{\infty} \beta(i, m) \frac{1}{N}\nonumber\\
		 \cdot \sum_{j=i N_L}^{i N_L+N-1}&\left\{\log \left({q}_{s, P D}(\mathbf{y}(j))\right)-\log \left({p}_{y, P D}(\mathbf{y}(j))\right)\right\},
	\end{align}

\noindent estimated over blocks of $N$ output samples shifted by $N_L$ relative to the previous block, $\beta$ is a window function expressed as a function of time sample ($i$) and block ($m$) indices, allowing for offline, online and block-online algorithms. The multivariate distribution of the multichannel signal at the separation system output is represented as ${p}_{y, P D}(\cdot)$, and the hypothesized source model ${q}_{s, P D}(\cdot)$ for blind source separation is often given as
\begin{equation}%
{q}_{s, P D}(\mathbf{y}(j)) = \prod_{p=1}^P {p}_{y_p, D}\left(\mathbf{y}_p(j)\right),
\end{equation}
with $D$ denoting the dimensionality of each individual multivariate source. Typically, non-Gaussian prototypical distributions are used for the density functions such as multichannel Laplacian with parameters estimated from the data. For the linear (over)-determined case, the output density in Eq.~(\ref{eq:trini}) can be expressed by means of the input sensor's signal probability density, which is independent of the parameters to be estimated, and the log determinant of the demixing system \cite{trini, hyvarinen2001independent}.


\section{Relation of Disentangled Variational Autoencoder to Blind Source Separation}
In this section, we will explore the relationship of Variational Auto-Encoders (VAEs) to the traditional BSS problem. VAEs are prominent deep neural network models that achieve both nonlinear dimensionality reduction and generative modeling. A VAE indirectly minimizes the Kullback-Leibler Divergence between a surrogate density for the latent space and the posterior actual latent space density by minimizing the following cost function, which can be interpreted as the negative of the evidence lower bound (ELBO) \cite{KingmaRMW14}:
\begin{align}\label{eq:vae}
	\mathcal{L}_r:= \mathcal{E}_{\mathbf{s}}\Big[\mathcal{E}_{\mathbf{z} \sim q_\phi(\mathbf{z} | \mathbf{s})}\left[-\log p_\eta(\mathbf{s} | \mathbf{z})\right]+&\operatorname{KL}\left(q_\phi(\mathbf{z} | \mathbf{s}) \| p(\mathbf{z})\right)\Big]\nonumber\\+ &\mathcal{D}\left(q_\phi(\mathbf{z}) \| p(\mathbf{z})\right),
\end{align}

\noindent with $\eta$ and $\phi$ 
denoting the parameters of the decoder and encoder respectively, and where $\mathbf{z}$ is the latent variable, $\mathbf{s}$ data sample. 
$\mathcal{D}$ is a regularization term that is often chosen to disentangle the latent space as proposed in the so-called DIP-VAE-II by promoting covariance matrices that are equal to the identity matrix \cite{disentangled}.

The variational autoencoder loss function and the classical blind source separation problem share many similarities. Recent research has focused on the possibility of finding disentangled data representations in an unsupervised manner, which has been shown to be theoretically not possible in the general case without inductive biases \cite{locatello2019challenging}. More recently, it has been shown that unsupervised disentangled representation is only possible in cases where certain conditions about the encoder and the latent representation are fulfilled, these conditions being non-Gaussianity and local isometry of the encoding scheme \cite{when}. For the task at hand of sound source separation, we are not interested in the 
classical goal of disentangled VAEs, which is a disentangled representation of a single sound source belonging to a certain manifold in order to enable generation of sound sources on this manifold by few explanatory factors. Rather, we are interested in finding a representation that disentangles sound sources in a mixture where the individual sound sources are described by means of multivariate probability densities. To further illustrate this motivation, consider the second order statistics based source separation approaches, e.g., \cite{trini} where the goal is to block-diagonalize the autocorrelation matrix of the output signals over multiple time lags rather than completely diagonalize it. Hence, we allow the latent representation of each source to have statistical dependencies but at the same time, we require them to be mutually independent. The inductive biases for our disentanglement task are defined by the ability to reconstruct the actual sound sources in the training dataset from the disentangled representation. Further, instead of requiring the latent space to be Gaussian as is common in VAEs, we explicitly enforce non-Gaussianity as this is a necessary condition for decomposing a mixture into independent components \cite{hyvarinen2001independent}.

Eq.~(\ref{eq:trini}) can be seen as a sample version of the Kullback-Leibler Divergence between the multivariate prior source distribution and the posterior. However, a maximum likelihood estimator of the blind source separation task would aim at maximizing the log likelihood of the sensor signals (mixture) given the latent variables. The chain rule allows the expression of the sensor signal log likelihood as 
\begin{align} 
	\log(p(\mathbf{x};\boldsymbol{\theta})) =\int q(\mathbf{z}) &\left[\log \left(\frac{p(\mathbf{x}, \mathbf{z} ; \boldsymbol{\theta})}{q(\mathbf{z})}\right)\right. \\\nonumber
	&\left. - \log \left(\frac{p(\mathbf{z} \mid \mathbf{x} ; \boldsymbol{\theta})}{q(\mathbf{z})}\right) \right] \mathrm{d} \mathbf{z},
\end{align}
with $\boldsymbol{\theta}$ being only the separation parameters, different from the encoder and decoder parameters ($\eta$, $\phi$). The first term in the integral is the ELBO as seen before in the VAE context. This can be written as
\begin{align}
\mathcal{L}_r=\mathcal{E}_{\mathbf{x}}\Big[\mathcal{E}_{\mathbf{z} \sim q_\phi}\left[-\log p_{\eta}(\mathbf{x} | \mathbf{z};\boldsymbol{\theta})\right]+\operatorname{KL}\left(q_\phi(\mathbf{z} | \mathbf{x};\boldsymbol{\theta}) \| p(\mathbf{z};\boldsymbol{\theta})\right)\Big],
\end{align}
where all of the terms can be computed for a suitable surrogate density $q_\phi$. This cost function leads to a VAE w.r.t. the mixture signal ($\mathbf{x}$). To enforce the disentanglement w.r.t. to the source, we use the following strategy:
First, instead of using the regularization term to promote disentanglement as in Eq.~(\ref{eq:vae}), we use as a surrogate, a distribution that promotes separation as is common in the classical BSS literature \cite{hyvarinen2001independent, Buchner2007}, e.g., a non-Gaussian multivariate spherically invariant random process distribution (SIRP).
Second, the reconstruction error as appearing in the ELBO is replaced by the term
\begin{align}\label{eq:bias}
\mathcal{D}_{\textrm{ind. bias}} = \sum_p \mathcal{E}_{\mathbf{z} \sim q_\phi}\left[-\log p_{\eta}(\mathbf{s}_p | \mathbf{A}_p\mathbf{z}; \boldsymbol{\theta})\right],
\end{align}
where $\mathbf{A}_p$ is a windowing matrix that selects the $p$-th block of the latent multivariate vector considered as a structured vector with $P$ blocks each of length $M$, and $\mathbf{s}_p$ represeting the ground truth separated source.%
 An interesting aspect of our formulation is that it uses the same decoder for the mixture as for the individual isolated sources, making
it inherently resource efficient.
For the purposes of this paper and for reasons described in the next section, we will be replacing the KLD minimization with the Fisher Divergence as a local approximation.


\section{Estimation by Score Matching}
In variational autoencoders, maximizing the ELBO entails minimizing the Kullback-Leibler Divergence between a prior latent distribution and an approximate posterior distribution assuming some tractable distribution. In the context of blind source separation, the prior over the latent variable is often given in unnormalized manner. For this purpose, a method called Score Matching (SM) has been developed where instead of matching the distributions, the derivative of the log densities are matched, which is a variational calculus approach to circumvent the need to know the normalization terms and leading to solutions robust to model misspecification  \cite{JMLR:v6:hyvarinen05a}. It can  be shown that SM minimizes the Fisher divergence between the data distribution and a model distribution which is specified by the gradient (with respect to the multivariate input variable $\boldsymbol{\zeta}$) of its log density function which is parameterized by $\boldsymbol{\xi}$, \cite{lyu2012interpretation, liu2019fisher}. %
SM introduces the score function as the derivative of the log density defined as
\begin{equation}
	\psi(\boldsymbol{\zeta} ; \boldsymbol{\xi})=\left(\begin{array}{c}
		\frac{\partial \log q(\boldsymbol{\zeta} ; \boldsymbol{\xi})}{\partial \zeta_1} \\
		\vdots \\
		\frac{\partial \log q(\boldsymbol{\zeta} ; \boldsymbol{\xi})}{\partial \zeta_{PM}}
	\end{array}\right)=\left(\begin{array}{c}
		\psi_1(\boldsymbol{\zeta} ; \boldsymbol{\xi}) \\
		\vdots \\
		\psi_{PM}(\boldsymbol{\zeta} ; \boldsymbol{\xi})
	\end{array}\right)=\nabla_{\boldsymbol{\zeta}} \log q(\boldsymbol{\zeta} ; \boldsymbol{\xi}) .\nonumber
\end{equation}
The cost function of SM reads
\begin{equation}
	J(\boldsymbol{\xi})=\frac{1}{2} \int_{\boldsymbol{\zeta} \in \mathbb{R}^{PM}} q_{\mathbf{z}}(\mathbf{\zeta})\left\|\boldsymbol{\psi}(\boldsymbol{\zeta} ; \boldsymbol{\xi})-\boldsymbol{\psi}_{\mathbf{z}}(\boldsymbol{\zeta})\right\|^2 d \boldsymbol{\zeta},\nonumber
\end{equation}
which has been shown in \cite{JMLR:v6:hyvarinen05a} to be equal to
\begin{equation}\label{eq:hyverinen}
	J(\boldsymbol{\xi})=\int_{\boldsymbol{\zeta} \in \mathbb{R}^{PM}} q_{\mathbf{z}}(\boldsymbol{\zeta}) \sum_{i=1}^{PM}\left[\partial_i \psi_i(\boldsymbol{\zeta} ; \boldsymbol{\xi})+\frac{1}{2} \psi_i(\boldsymbol{\zeta} ; \boldsymbol{\xi})^2\right] d \boldsymbol{\zeta}+\text { const. }
\end{equation}
where
\begin{equation}
	\psi_i(\boldsymbol{\zeta} ; \boldsymbol{\xi})=\frac{\partial \log q(\boldsymbol{\zeta} ; \boldsymbol{\xi})}{\partial \zeta_i}\quad\text{, and}\nonumber
\end{equation}
\begin{equation}
	\partial_i \psi_i(\boldsymbol{\zeta} ; \boldsymbol{\xi})=\frac{\partial \psi_i(\boldsymbol{\zeta} ; \boldsymbol{\xi})}{\partial \zeta_i}=\frac{\partial^2 \log q(\boldsymbol{\zeta} ; \boldsymbol{\xi})}{\partial \zeta_i^2}.\nonumber
\end{equation}

\noindent This cost function promotes realization of the log density that fits the data while balancing the curvature and the steepness. When the gradient norm is zero, the data point is at a local extremum, and the Laplacian determines the curvature of the log likelihood; if negative, the curvature is convex and the extremum is a maximum. %
\noindent In this study, we are interested in non-Gaussian distributions motivated by the condition for the separation into independent components \cite{hyvarinen2001independent}. Motivated by the choice in \cite{trini}, we select SIRP as a family of non-Gaussian distributions. A prominent realization of this family is the multivariate Laplacian, for which the score function is \cite{Buchner2007}
\begin{equation}
	\psi\left(\mathbf{z}_p; \operatorname{C}_{\mathbf{z}_p}\right)=-2\cdot\frac{1}{\sqrt{2 u_p}} \frac{K_{M / 2}\left(\sqrt{2 u_p}\right)}{K_{M / 2-1}\left(\sqrt{2 u_p}\right)}{\operatorname{C}^{-1}_{\mathbf{z}_p}} \mathbf{z}_p,
\end{equation}
with $K_n$ denoting the n-th order modified Bessel functions of the second kind, and 
\begin{equation}
	u_p=\mathbf{z}_p^{\mathrm{T}} {\operatorname{C}^{-1}_{\mathbf{z}_p}} \mathbf{z}_p,
\end{equation}
with $\operatorname{C}_{\mathbf{z}_p}$ representing an estimation of the covariance matrix of one estimated separated source in the latent space.


\vspace{-.1cm}
\section{Differentiable Permutation Minimization}\vspace{-.2cm}

In source separation, there are two common permutation problems:  source permutation, in general, if the network has no further conditions on the statistics of each output, the order of the output streams is random; and frequency permutation where the order of the separated sources is inconsistent across different frequency bands. 
A recurrent architecture can help mitigate the issue as it helps keeping context of the separated sources. However, for extended periods of silence, the problem could occur further. Adding dedicated memory layers on some representation of the embedding or longer sequence information has been suggested in \cite{luo2020dualpath, li2022skim} to circumvent this issue. 
In the present work, we introduce a way to incorporate permutation mitigation into the loss function, further, augment the architecture with a memory function to keep track of the individual source statistics.
The negative of the losses for the different permutations in a permutation invariant training \cite{yu2017permutation} can be seen as unnormalized logits of a categorical distribution. The permutation would be the argmax of the logits which is not differentiable, instead, we use the softmax across the output channels. An undesirable permutation would occur if the network  exchange the sources across frames or subbands. In order to minimize such permutations we constrain the derivative of the differentiable permutation at multiple resolutions. A multiresolutional differential operator can be obtained as the second to last columns of the Hadamard matrix defined for $i\in \mathbb N$ as
\begin{equation}
H_{n=2^{i}}=	\left[\begin{array}{cc}
		H_{2^{i-1}} & H_{2^{i-1}} \\
		H_{2^{i-1}} & -H_{2^{i-1}}
	\end{array}\right]. 
\end{equation}
A loss function minimizing the permutation across subbands is established based on the norm of the softmax output multiplied with the 2$^{nd}$ to last columns of the Hadamard matrix.
The permutation across frames is minimized by considering the structured norm $\ell_{1,2}$ of the discrete derivative of the softmax output across frames for each subband, where the $\ell_1$ is used across frames and then an $\ell_2$ is taken across frequencies over the previously calculated $\ell_1$ norms. 

\subsection{Memory Block Aided Permutation Solution}
The idea of augmenting sequence-to-sequence models with memory is well known, e.g., \cite{wu2022memformer}, for source separation the employment of memory block has been recently introduced with success in \cite{li2022skim}. To robustify our model against permutation errors in the presence of longer sequences of silence of the individual sources, we augment the network with a structured memory bank keeping track of the separated signal latent representations. This information is fed back to the internal separation as conditioning input. The latent representation of the output is buffered into a ring buffer with a length $K$ and an adaptive average pooling is performed over the buffered signals, the pooling output is fed into a gated recurrent layer serving as memory, the output of the recurrent layer is fed along with the internal separator input into a fully connected layer to form the new latent separation unit input. Each separated source latent representation has its own memory. Since the model includes a feedback loop from the output into the memory and back into the latent space separator unit, see Fig.~\ref{fig:sys}, during training we forward pass the signal with frozen parameters to obtain the the memory input and then train with a copy of the memory output detached from the computational graph as highlighted in the figure by the dashed arrow. The memory unit is trained in an alternating manner with rest of the network by incorporating the memory output within the graph into a contrastive predictive loss function designed in a manner similar to the contrastive predictive coding concept \cite{oord2019representation}. To obtain the targets for this loss, the clean separated ground truth sources are passed through the network with frozen parameters except those of the memory unit bypassing the fully connected layer which incorporates the memory input (see the dashed line skipping the layer before the separation unit). The output of the latent space separation unit is divided across the time axis into segments with the same length as the buffer length. The loss is similar to InfoNCE loss \cite{oord2019representation} with positive samples being frames from the following time segment of the same source and negative samples being from the following segments of other sources in the batch sequences 
\begin{equation}
\mathcal{L}_{\mathrm{N}}=-\underset{Z}{\mathbb{E}}\left[\log \frac{f\left(\mathbf{z}_{t+k}, \mathbf{c}_t\right)}{\sum_{\mathbf{z}_j \in Z} f\left(\mathbf{z}_j, \mathbf{c}_t\right)}\right],
\end{equation}
where $f(\mathbf{e},\mathbf{g})$ = $\exp(w\cdot \cos({\mathbf{e}},{\mathbf{g}}) + b)$ with trainable $w$, and $b$, and $\mathbf{c}_t$ denoting the memory content at a time $t$ and $k$ parameterizes the time horizon of positve samples obtained from the same source.


\section{Architecture}
The overall architecture of the proposed system is presented in Fig.~\ref{fig:sys}. 
The encoder in the model includes recurrent layers (GRUs) to account for long-term dependencies over the feature map.  The encoder transforms the input signal into a high dimensional representation 
in the latent space that can be considered after the concatenation layer in the diagram as analogous to
a virtual microphone array signal. To emulate a non-under-determined source separation case, the number of virtual microphones in the latent space is chosen to be larger or equal the number of the desired sources $P$.
The latent space signal branches into a mask calculation unit consisting of a convolutional layer and a sigmoid activation layer. A skip connection applies the mask on the ``virtual microphone" signals. The output is fed to a fully connected layer and then a decoder, which decodes one isolated signal at a time since all output sources share the same decoder. 
In the proposed network, the separation in the latent space is performed via a multiplication with an adaptive matrix ($\mathbf{z} = \boldsymbol{\theta}\mathbf{z^\prime}$), \textit{emulating} the (over)-determined multichannel separation of $P$-sources ($\mathbf{z} = [\mathbf{z}_1,\ldots,\mathbf{z}_P]^\mathrm{T}$) given $Q$-microphones ($\mathbf{z^\prime} = [\mathbf{z^\prime}_1,\ldots,\mathbf{z^\prime}_Q]^\mathrm{T}$)
where  $\boldsymbol{\theta}$ has the structure
\begin{equation}
\boldsymbol{\theta}=	\left[\begin{array}{c}
		\mathbf{F}\cdot \boldsymbol{\theta}_1 \\
		\ldots\\
		\mathbf{F}\cdot\boldsymbol{\theta}_Q
		\end{array}\right],
\end{equation}

with $\mathbf{F}$ being a matrix fixed after training with the dimensions $L_{out}\times L_{in}Q$. 
Since it is an-element-wise multiplication followed by a fully connected layer, we are encouraging the network to take the latent space interdependencies into account, similar to the solutions of classical DSP approaches to \textit{multivariate} BSS with the difference that now the separation itself is done in a latent space.
\begin{figure*}[!h]
	\centering
	\includegraphics[width=.85\linewidth]{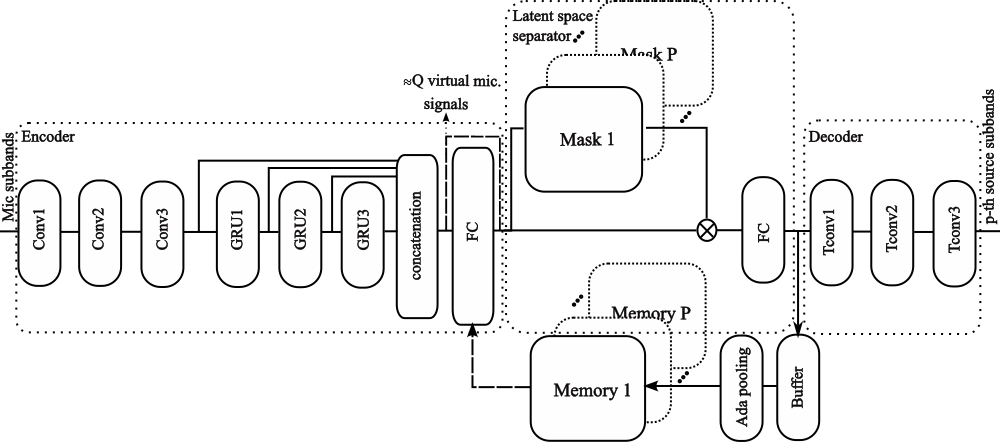}\vspace{-.3cm}
	\caption{Latent space disentangling blind source separation system architecture.}
	\label{fig:sys}\vspace{-.5cm}
\end{figure*}
\section{Loss function}
The loss function in our permutation-invariant training strategy
includes multiple components. We use the disentanglement loss from Eq.~(\ref{eq:hyverinen}) and additionally incorporate a subband-based SI-SDR \cite{roux2018sdr} and $\ell_{2}$ reconstruction loss as a realization of the inductive bias constraint in Eq.~(\ref{eq:bias}), and the mentioned permutation and memory bank losses.
\section{Experimental results}
	\label{sec:results}
	To prove our concept we train a model based on the proposed architecture with an encoder consisting of 3 convolutional layers with kernels of 1, 3, and 5, 3 recurrent sections each consisting of 3 GRU layers with skip between the sections connections and hidden size of 200, a fully connected layer and a decoder with 3 transposed convolutional layers. The FFT size is 1024, 
	the window size is 512, and the hop size is 256 from which we calculate 100 Bark band features which constitutes the network processing domain meaning that the network output is in the Bark domain but for calculating the losses, the signal is transformed first back into the full subband resolution.  
	For the memory, we use an 2 LSTM layers for each source to track the adaptive pooling output operating on a buffer storing the last 50 frames of the separated latent representation.
	We train the network on an in-house acquired dataset with explicit recording consent. For training we use mixtures of two clean fully overlapped speech signals without applying any further augmentation and train to output the separated mixture components. For testing, we use held out mixtures of unseen speakers with and without background noise. %
	The clean and noisy conditions have normally distributed SNRs with means of 0dB, -2dB and standard deviations of 3dB and 3.8dB respectively.   
	For a comparison, we train SkiM \cite{li2022skim} model consisting of 4 SkiM blocks. In each SkiM block,the LSTMs’s hidden dimension is 256. The segment LSTMs are chosen to be bi-directional and the segment sequence length is set to 150, the MemLSTM are uni-directional, and the feature normalization is performed on the feature dimension. Although the literature reports higher state-of-the-art numbers on synthetic datasets, in our experiment the test dataset is chosen to cover real-world scenarios with different conditions, different speakers and speaker styles than the training set.
	The results shown in Table \ref{table} confirm that the network generalizes for the noisy case while performing consistently on the clean data.

\begin{table}[!h]\centering
\begin{tabular}{c|c|c}
	\hline & \multicolumn{2}{c}{ \textbf {SI-SDR} } \\
	\hline & \textbf {Proposed} & \textbf {SkiM} \\
	\hline Noisy &   5.3 & 2.8\\
	\hline
	\hline Clean &  7.2 & 7.1\\
	\hline
\end{tabular}
\caption{SI-SDR achieved under noisy and clean condition for the online system trained on only clean mixtures.}\label{table}
\end{table}

\section{Conclusions}
In this paper, we presented an online approach to source separation motivated from first principles highlighting the links to classical DSP solutions to source separation. The method aims at a block-wise disentanglement in a latent space. 
We account for the permutation problem and long-term dependencies by using per-source memory augmentation and 
a novel differentiable permutation loss.
	\bibliographystyle{IEEEbib}
	\bibliography{refs}
	\end{sloppy}
\end{document}